# Sub-wavelength mid-infrared imaging of locally driven photocurrents using diamond campanile probes

**Rajasekhar Medapalli**[1,*], **Nathan D. Cottam**[2], **Khushboo Agarwal**[1,3], **Benjamin T. Dewes**[2], **Nils Dessmann**[4], **Sergio Gonzalez-Munoz**[1], **Wenjing Yan**[3], **Vaidotas Mišeikis**[5], **Sergey Kafanov**[1], **Rostislav V. Mikhaylovskiy**[1], **Samuel P. Jarvis**[1], **Camilla Coletti**[5], **Britta Redlich**[4], **Amalia Patanè**[2] and **Oleg V. Kolosov**[1,*]

[1]Department of Physics, Lancaster University, Lancaster LA1 4YB, UK
[2]School of Physics and Astronomy, University of Nottingham, Nottingham NG7 2RD, UK
[3]Department of Physics, School of Natural Sciences, Shiv Nadar Institution of Eminence (Delhi-NCR), Dadri UP 201314, India
[4]HFML-FELIX laboratory, Radboud University, Toernooiveld 7, 6525 ED Nijmegen, The Netherlands
[5]Center for Nanotechnology Innovation @ NEST, Istituto Italiano di Technologia, Piazza San Silvestro 12, 56127 Pisa, Italy

**\*Corresponding authors**
Email: <rajasekhar.medapalli@gmail.com>; <o.kolosov@lancaster.ac.uk>

## Abstract
Precise and high efficiency concentration of mid-infrared (mid-IR) light into sub-wavelength volumes is essential for probing low-energy excitations and achieving strong field enhancements, which can be hindered by absorption losses and coupling inefficiencies at long wavelengths. Here, we introduce an innovative diamond-based metal-insulator-metal campanile probe that adiabatically compresses free-space mid-infrared light (10 μm) into ≈1 μm domains. Integrated into a scanning photovoltage microscope, the probe enables sub-wavelength mapping of locally driven photocurrents in graphene, resolving polarization dependent and contact-sensitive responses at energies down to ≈0.1 eV. Experiments reveal a photocurrent signal density enhancement of $10^3$ and coupling efficiencies approaching 80%, in agreement with numerical simulations. Operation of the probe with quantum cascade and free-electron lasers demonstrates a robust, spectrally tunable platform for high-resolution exploration of low-energy carrier dynamics in atomically thin materials, opening opportunities for mid-IR optoelectronics and quantum photonics.

## Introduction
Mid-IR light offers a powerful insight into low-energy electronic and vibrational excitations, providing access to nonequilibrium phenomena[1-5] that remain elusive to conventional optical techniques. Owing to their reduced photon energy compared to optical photons, intense mid-IR fields selectively couple to low-energy modes, driving nonthermal carrier redistribution[6], hot photocurrents[7], and transient symmetry breaking[8]. When confined to the nanoscale, these electric fields exhibit enhanced intensities that unlock otherwise inaccessible strongly nonlinear and ultrafast responses. Capturing these effects demands sub-wavelength mid-IR imaging, which enhances light-matter coupling[9,10] and directly probes the mechanisms governed by the strength of electromagnetic interaction between quasiparticles, especially in two-dimensional materials, such as graphene[11,12]. Achieving spatial resolution down to intrinsic diffusion lengths of charge and heat carriers[7] is particularly valuable for directly interrogating photocurrent generation and transport pathways, providing critical insights into the microscopic nature of nonequilibrium charge dynamics. Combined together, these capabilities reveal light-matter interactions, extending equilibrium and semiclassical frameworks[13-15], and open new pathways to disentangle the microscopic mechanisms of light-induced energy flow and dissipation in low-energy quantum systems[16]. This particularly applies to the case when the same probe simultaneously concentrates fields for spatial localization and provides direct access to mid-IR driven photocurrent generation.

Adiabatic field compression using campanile-shaped probes offers a broadband, non-resonant, and highly efficient alternative to conventional aperture scanning near-field optical microscopy (SNOM) or scattering (apertureless) type S-SNOM. It delivers a significant fraction (on the order of 50% compared to $10^{-3}$-$10^{-4}$ in aperture SNOM) of the input light exclusively to the area of interest at micro/nano length scales, without parallel illumination of the adjacent areas (as in apertureless S-SNOM), and collects the light that interacted with the object from the same area, also with high efficiency. However, in the mid-IR regime optical components suffer from intrinsic challenges, including increased absorption losses originating from free-carrier damping in noble metals or phonon-polariton absorption in dielectrics materials typically employed in far-field to near-field transformers. Here, we introduce a custom-fabricated metal-insulator-metal (MIM) campanile diamond probe, optimised for mid-IR operation and integrated into a commercial scanning near-field microscopy platform, achieving 80% coupling efficiency into sub-wavelength focus at the apex, in good agreement with numerical simulations. The probe features a tapered tetragonal-pyramid geometry consisting of a dielectric core flanked by two opposing metal-coated facets, while the other two sides remain uncoated (Fig. 1a-c). It enables smooth, broadband funnelling of free-space radiation with minimal reflection or scattering. This non-resonant principle allows strong field enhancement without the spectral limitations and dissipative losses inherent to plasmonic or aperture-based designs. While related concepts have been investigated at visible and near-IR spectral ranges[17-20], our approach uniquely covers adiabatic compression into the mid-IR with potential of expanding in THz regime, demonstrating broadband compatibility (8.5 - 10.5 µm) and robustness with high power, pulsed free-electron laser (FEL), capabilities unattainable with traditional SNOM techniques.

We leverage this platform to spatially map photocurrents proportional to photovoltaic (PV) signals in patterned, gate-tunable graphene devices, revealing directly polarization-dependent and contact-sensitive responses. The Au/diamond/Au campanile probe enables deep mid-IR light confinement with field enhancement by 3 orders of magnitude, achieving ≈1 µm spatial resolution and allowing direct imaging of micron-scale photocurrent distributions near junctions and contacts. This sets a new benchmark for mid-IR nanoscale spectroscopy, especially when apex dimensions are further reduced, as predicted by finite-difference time-domain (FDTD) simulations. By combining adiabatic compression with spatially resolved photocurrent mapping, we establish a broadly applicable framework to study lower-energy carrier contributions to photo-thermoelectric effects, offering a versatile and tunable platform for investigating complex quantum systems.

**Results**

Surface plasmon polaritons (SPPs) are formed at a metal-dielectric interface due to the coupling of incident photons with the collective oscillations of free electrons in the metal. In campanile pyramids, SPP generation spans a broad spectral range, dictated more by structural geometry and the broadband properties of dielectric and metal, rather than by the intrinsic plasmonic properties of the materials. For wavelengths larger than the lateral gap between the facets of pyramid, defined as *a*, momentum matching conditions are satisfied, enabling adiabatic concentration of light at the apex[21-25]. Unlike conventional plasmonic antennas with fixed resonances, where coupling occurs at a fixed resonance set by material permittivity and geometry, the campanile design supports a continuum of modes through gradual tapering, allowing broadband light concentration. Efficient coupling occurs when the base width exceeds the wavelength, transforming far-field light into near-field modes that travel along the dielectric/metal interfaces and converge at the tip. Theoretical work by Polyakov et al.,[26] predicts that inverted insulator-metal-insulator (IMI) pyramids can sustain strong apex field enhancement for wavelengths well beyond the near-IR, up to $\approx$10 µm and beyond provided dielectric transparency at the wavelengths used.

To validate our hypothesis, we fabricated a diamond-based MIM campanile pyramid (Fig. 1c) and conducted 3D finite difference time domain (FDTD) simulations using ANSYS Lumerical software to model its properties[18,19,26]. The simulations were targeting mid-IR light parameters with 100 fs pulse duration and a central wavelength ($\lambda$) of 10 µm. To optimize the geometry for enhanced mid-IR focusing at the apex, we systematically varied the apex dimensions (along x and y axes) and the pyramid apex-angle ($\theta$) (see Supplementary Figs. 1 and 2). Frequency domain field monitors and time monitors were strategically placed to capture cross-sections of the steady state and time-varying electric field (E-field) behavior at the apex, and along the xy and yz planes within the pyramid during the adiabatic compression process (see Supplementary Fig. 1).

The simulations in Fig. 1d reveal that the chosen geometry efficiently couples mid-IR far field to the near-field domain. For an apex size of 2000 nm × 100 nm and $\theta = 40°$, an adiabatic spot size is achieved with nearly zero background. The strong *E*-field enhancement at the diamond/metal interfaces confirms that the mid-IR transformation from far-field to near-field is mediated by SPPs (wavevector $S_0$). This is further confirmed by the apex size dependence, where the *E*-field enhancement exhibits a strong increase as the apex size shrinks to 10 nm × 10 nm. The data points in Supplementary Fig. 2d correspond to various apex sizes, with specific dimensions highlighted in the inset. A campanile pyramid was fabricated on a chemical vapor deposition (CVD) grown diamond using gold metallized facets, as illustrated in Fig. 1b, producing a Au/diamond/Au campanile pyramid with an apex size of 3 µm × 100 nm (with the longer dimension parallel to the probe prismatic edge and the metallized facets) and $\theta = 80°$. To evaluate the functionality of our mid-IR transformer, we used

CVD grown patterned graphene-devices. For our experiments we considered two different types of devices: graphene-channel capped with a flake of the 2D material GaSe and uncapped graphene. The optical images of the graphene device are shown in Fig. 2a. The PV response of these two different types of devices in the far-field configuration showed nearly identical behavior (Supplementary Fig. 3). The mid-IR PV response of the graphene was measured to characterize the adiabatically concentrated spot size and probe localized PV response, demonstrating the effectiveness of the mid-IR transformer.

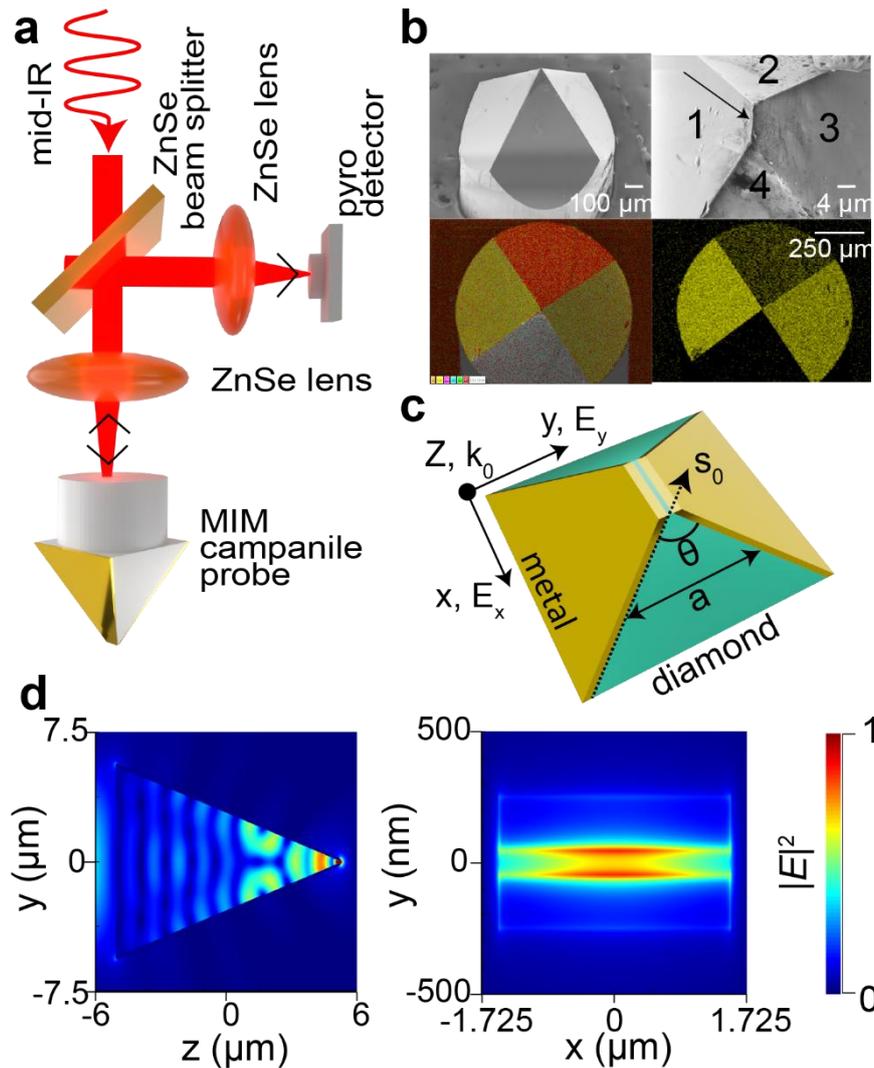

**Fig.1 | Metal-insulator-metal (MIM) campanile light transformer and its principle of operation.** (a) Schematic of the near-field mid-IR laser excitation setup with simultaneous detection of PV response and reflected light signals. For far-field measurements, the MIM campanile probe was retracted allowing direct sample illumination with a larger spot size (≈30 μm). The laser polarization was set either to $E_x$ or $E_y$ using a combination of half-wave plate and polarizer (not shown). (b) top panel: Scanning electron microscopy (SEM) images illustrating the structure of the campanile shaped diamond pyramid. The numbers in the right-side image identify the facets of the pyramid, with the arrow highlighting the prismatic line at the apex. Facets labelled "1" and "3" are metallized to enable the generation of an adiabatic light spot at the apex. Bottom panel: Electron dispersive x-ray spectroscopy (EDS) maps showing the chemical composition of the metallized pyramid. The maps show the distribution of carbon in orange (left), Cu-kα1 in yellow (right), confirming the materials used in the fabrication process and selective metallization of the facets 1 and 3. (c) Schematic of a MIM campanile shaped tetragonal pyramid. (d) Finite element simulation showing (left) the yz cross-section of far field to near field conversion of mid-IR light (≈10 μm) within the campanile pyramid while focusing the light, (right) background-free near-field enhancement at the apex of a metal/diamond/metal campanile pyramid, where the apex size is set to 3 μm × 100 nm, and the metal thickness on each side of 200 nm.

First, we examined the mid-IR far-field PV response of the graphene device using a photovoltaic measurement scheme as illustrated in Fig. 2b and with the device excited by a pulse train from a quantum cascade laser (QCL, $\lambda$ = 9.5 μm). For reference far-field measurements, the MIM campanile probe was removed allowing direct sample illumination with a large spot size (≈30 μm). This test excluded the campanile probe and instead used a ZnSe lens to focus light. Figures 2c and 2d present area mappings of simultaneously measured reflectivity and photovoltage ($V_{PV}$) response, respectively, in the far-field configuration. These maps were obtained by translating the lens parallel to the sample plane to focus the laser beam onto different regions of the sample surface. The intensity map provides an overview of the device's reflectivity, identifying key regions of interest such as the gold contacts and the graphene-layer. The background-free PV signal measurement reveals two distinct spots (Fig. 2d), indicating maximum PV obtained when the beam is centred at the gold contacts. The opposite sign of the PV response at the two contacts confirms the expected polarity dependence.

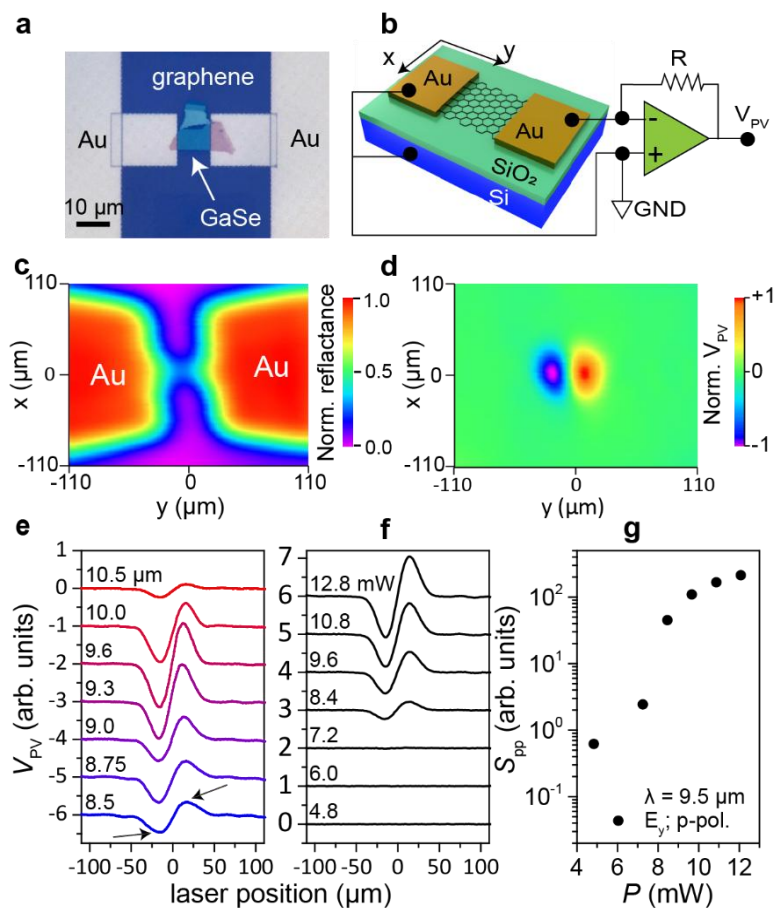

**Fig.2 | Far-field reference of mid-IR PV response.** (a) Optical micrograph of a Au-graphene-Au planar device capped with a GaSe flake. (b) Schematic illustration of the graphene device and photovoltage measurement configuration. The Au/graphene interfaces are along the y-axis. (c,d) Spatial map of the reflected light intensity (c) and PV signal (d) as a 9.5 μm light beam with a spot size of ≈30 μm is scanned across the GaSe/graphene surface. (e) Measured PV response for different excitation $\lambda$ and (f) for different $P$ at $\lambda$ = 9.5 $\mu$m. For each $\lambda$, the maximum output power available from the QCL was used. The arrows indicate the Au-electrode edges. The spectral dependence of the maximum power, $P_{max}$ at the sample plane is shown in Supplementary Fig. 4. (g) The extracted $S_{pp}$ values as a function of $P$ ($\lambda$ = 9.5 $\mu$m).

To probe spatial variations, line profiles were obtained for wavelengths ranging between 8.5-10.5 μm by moving the laser beam across a distance of 220 μm, covering the graphene/metal junctions, as shown in Fig. 2e. The profiles reveal that the $V_{PV}$ is maximized when the laser beam is positioned at either of the contacts, with opposite polarities observed at the left and right Au-contact regions. In contrast, at the center of the sample, the PV signal goes to zero. Figure 2f shows the excitation laser power

(P) dependence at $\lambda$ = 9.5 µm. The strength of the PV response $S_{pp}$ (the peak-to-peak value between the left and right peaks) is plotted against $P$ in Fig. 2g. Above a threshold power, the PV signal increases with $P$ and saturates at $P$ = 12 mW, indicating a nonlinear response.

The mid-IR near-field response of the graphene-device is presented in Fig. 3. In this case, the Au/diamond/Au campanile probe with apex dimensions of 100 nm x 4 µm is used both to focus the incident light into a highly confined slit as well as to collect the reflected light (see Fig. 1a). The device's topography, reflected light intensity, and $V_{PV}$, were acquired simultaneously using the Au/diamond/Au campanile probe and are shown in Figs. 3a, 3b, and 3c, respectively. The polarization of the incoming light and power are set to $E_y$ and 12 mW, respectively. The reflectivity data obtained using a near-field scan over the surface of the Au/graphene (left-)interface is plotted in the inset of Fig. 3b. The high and low intensity values correspond to Au and graphene sample regions, respectively. The transition between these regions, observed as a slope in the intensity curve, is indicative of the spot size of the light focused at the pyramid apex. By fitting the data shown in the inset of Fig. 3b with an error function (Erf), a spatial resolution of 1.01 ± 0.06 µm is obtained, approximately an order of magnitude smaller than the excitation wavelength ($\lambda$ = 9.5 µm). The fitted curve is overlaid in red on the plot, providing compelling evidence for the successful far-field to near-field transformation achieved using the MIM campanile pyramid.

Similarly, horizontal line profiles of the $V_{PV}$ signal, obtained as the MIM probe scans across the device surface are shown in Figs. 3d for two orthogonal polarizations of the laser ($E_y$ and $E_x$, respectively). The data where polarity of the device is inverted, leading to a reversed sign in the PV signal (see Supplementary Fig. 5), indicate a photovoltaic effect, rather than a photoconductive response. The PV signal reveals multiple peaks, distinct from those observed in the far-field configuration, which are confined to the Au/graphene interfaces. For comparison, the corresponding far-field data is overlaid in blue in Fig. 3d indicating a lower spatial resolution. The observed fine features in the photoresponse of the device highlight the capability of sub-wavelength focusing of mid-IR light by the MIM campanile probe, enabling localized probing and manipulation of the device spatial response. Most importantly, when the laser polarization is switched to $E_x$, a dramatic difference emerges at the center of the device (probe position at 0 µm): the PV signal becomes significantly stronger at the center while weakening near the Au/graphene interface. This behavior is in striking contrast to $E_y$ excitation, while the positions of the additional features remain unchanged.

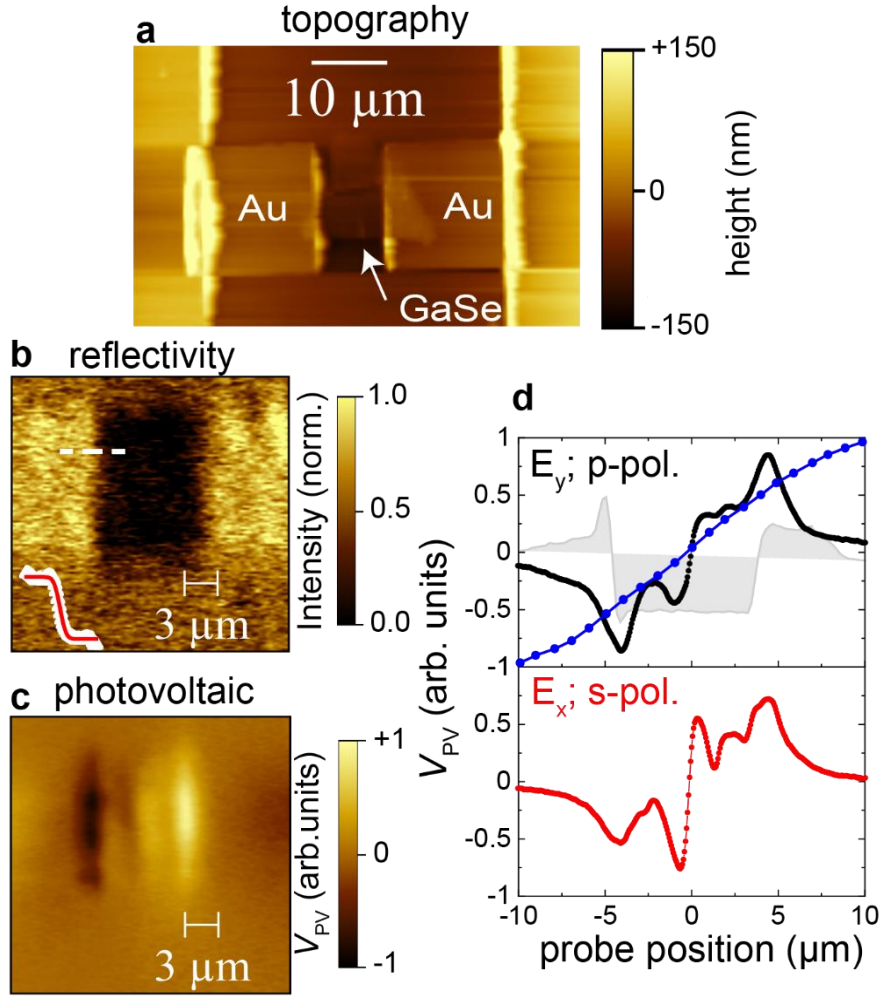

**Fig.3 | Near-field mid-IR photovoltaic response.** (a) Topography image of the graphene device with gold contacts for electrical measurements obtained in the SPM mode using the MIM diamond probe as the apex scanning tip. The graphene-layer is capped with GaSe, marked by a white arrow. Simultaneously measured reflectivity (b) and PV response (c) while scanning the MIM probe across the device surface, focusing on the region Au-graphene/GaSe-Au region. The reflectivity drop between the gold and sample regions is shown in the inset of (b) for a horizontal length of 5 μm (white dashed line). The red curve is the error function fit yielding a spatial resolution of 1.01 ± 0.06 μm. (d) Horizontal line profiles of the PV signal for two orthogonal excitation laser polarizations. The grey shaded curve shows the corresponding horizontal line profile of height. The blue curve shows the far-field data with a lower spatial resolution. The arrows indicate the Au-electrode edges.

The observed mid-IR $V_{PV}$ signal in the graphene-device is strongly localized near the Au/graphene contacts and exhibits a threshold-like nonlinear dependence on $P$, indicating a thermally driven process. Since the mid-IR photon energies are too low (130 meV at 9.5 μm) to drive efficient interband transitions but are well suited to heating the electronic system via intraband (Drude-like) absorption, the PV signal is dominated by thermal effects rather than direct photovoltaic mechanisms[12,27,28]. This is consistent with the strong PV response arising from electronic heating and subsequent thermoelectric conversion. As shown in Fig. 3d, the PV signal also exhibits a pronounced dependence on the polarization of the incident light. Under $E_y$ polarization, where the $E$-field is aligned parallel to the Au/graphene interface, the signal is enhanced at the contacts. In contrast, for the orthogonal polarization $E_x$ (perpendicular to the interface), a stronger PV signal is achieved within the graphene channel. This polarization dependent behaviour points to anisotropic absorption mechanisms, such as the polarization-selective excitation of near-field or plasmonic modes near metal edges[29-34], which can concentrate mid-IR light energy, thereby enhancing local heating effects. The resulting spatially localized temperature gradients drive photovoltages via the photo-thermoelectric effect, particularly in the presence of spatially varying Seebeck coefficients potentially due to contact-induced doping gradients[27,35]. Furthermore, the photovoltage exhibits a smooth sign reversal as a function of the gate voltage (Supplementary Fig. 6), consistent with a bipolar Seebeck response that crosses the Dirac point. This gate tunable behaviour, combined with mid-IR thermal excitation and polarization-dependent absorption[2], strongly supports that the dominant low-energy driven photocurrent generation mechanism in our graphene-device under mid-IR illuminations arises from the photo-thermoelectric effect.

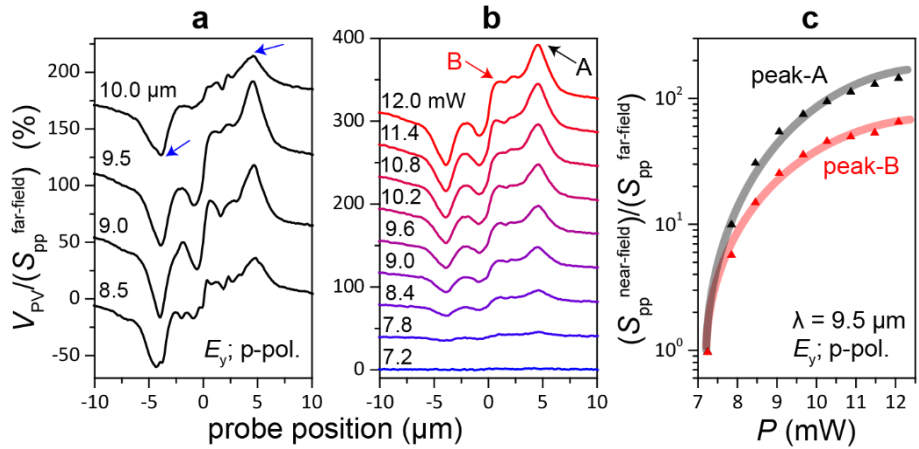

**Fig.4 | Spectral and power dependence of near-field mid-IR photovoltaic response.** (a) Line profiles of the $V_{PV}$ response measured at four different $\lambda$, as the probe is scanned across the center of the device for a horizontal distance of 20 μm from left to right. The graphene-layer is capped with a GaSe flake. The blue arrows indicate the Au-electrode edges. (b) Line profiles of the $V_{PV}$ response measured for various laser powers ranging from 7.2 to 12.8 mW, with $\lambda$ fixed at 9.5 μm and $E$-field set to $E_y$. (c) The normalized $|S_{pp}|$ values as a function of $P$ for two different peaks as indicated by A (black triangles) and B (red triangles) in part (b). The solid curves are guide to the eye.

Figs. 4a and 4b present the line profiles of the spatially resolved near-field $V_{PV}$ signal across the graphene device, as a function of $\lambda$ and $P$, respectively, with values normalized to the $|S_{PP}|$ value obtained for the far-field measurement case ($\lambda$ = 9.5 um, $P$ = 12.8 mW, $E_y$-polarization). At shorter wavelengths ($\lambda$ = 8.5 μm), the PV signal exhibits secondary peaks beyond the primary ones near the Au contact regions. These weaker, spatially resolved features vary with excitation wavelength (see Fig. 4a), indicating a sensitivity to the spectral characteristics of the incident light. We attribute this to the wavelength-dependent spatial absorption modulation within the graphene layer, arising from interference-enhanced absorption in the Si/SiO$_2$ substrate or coupling to substrate phonon-polariton modes[12,36,37]. Such effects can lead to localized variations in electron temperature gradients, especially in the mid-IR spectral regime where graphene's finite optical conductivity and the substrate's dielectric contrast result in non-uniform absorption[36,38,39]. These localized thermal gradients, when combined with spatially varying Seebeck coefficients, generate spatially distinct PV signals consistent with the photothermal mechanism. The emergence of these distinct, wavelength-dependent features within the graphene-channel, away from the metal contacts, is observable only by using the MIM campanile probe. The spatial resolution achieved was approximately 1 μm, this represents a nearly 30-fold improvement over the ≈30 μm spot size used in the far-field geometry. This resolution enhancement enables mapping of the photovoltaic response, as it would be spatially averaged out in far-field measurements. The near-field configuration allowed us to concentrate mid-IR energy into micrometer-scale regions with high precision, thereby selectively exciting local photothermal gradients. This capability provides access to spatially distinct photo-thermoelectric responses, revealing features at reduced length scale, such as polarization and wavelength dependent absorption within the graphene active layer that cannot be resolved in the far field.

Figure 4c highlights the $E_y$ polarization, normalized $|S_{PP}|$ values at the contact and within the channel region, as indicated by A and B, respectively in Fig. 4b. While the absolute signal levels differ, both curves exhibit a similarly nonlinear dependence on $P$. This consistent trend across distinct spatial regions suggests a shared underlying photothermal mechanism driving the observed response.

In the $\lambda$-dependence of $V_{PV}$ signal shown in Fig. 4a, it is evident that the near-field $V_{PV}$ signal ($\lambda$ = 9.5 µm) reaches up to ≈80% of the far-field maximum demonstrating that most of the incident mid-IR energy is effectively concentrated into a micrometer-scale region, primarily near the Au/graphene interfaces where localized photothermal gradients are strongest for $E_y$-polarization. This indicates that, despite reflection losses at the air/diamond interface and extreme sub-wavelength confinement, most usable light energy is preserved and efficiently delivered to the probe apex. FDTD simulations suggest an $E$-field intensity enhancement of ≈$10^4$ at the probe apex of the MIM probe (see Supplementary Fig. 2d). However, the experimentally observed enhancement in signal density (nearly 850 times increase per unit area) is somewhat lower, likely due to practical limitations such as tip-sample spacing and sample absorption losses. Nonetheless, the near-field configuration yields a signal comparable to the far-field response while probing nearly three orders of magnitude smaller area, confirming substantial local field concentration and effective energy delivery. While maximum photoresponse in the graphene-channel is expected for $E_x$-polarization due to its alignment with the channel axis, the observed nonlinear power dependence under $E_Y$ polarization is still consistent with thermoelectric mechanisms. Future measurements under $E_X$ will help disentangle anisotropic response contributions more precisely.

We assessed the operation of the probe with a free electron laser (FEL) which has a pulse operation with a very different time signature from that of QCL (see Supplementary Fig. 7). Figure 5a shows the topography (top panel) and PV response (bottom panel) maps of a graphene device as imaged using the mid-IR probe and the FEL as the light source at 10.5 µm. Horizontal line profiles, extracted from the center of the images, are plotted in Fig. 5b, providing a direct comparison between the spatial variations in the device's topography and its localized photovoltaic behavior. Building on our simulations, the area-averaged field intensity enhancement was found to scale nearly inversely with the apex area ($<|E|^2>/|E_0|^2 \propto A^{-0.95}$), revealing that further miniaturization of the focusing region can substantially increase the optical confinement. Extrapolation to apex dimensions approaching 10 nm x 10 nm suggests the possibility of achieving orders of magnitude higher near-field intensities in the mid-IR spectral range, while maintaining adiabatic energy transfer.

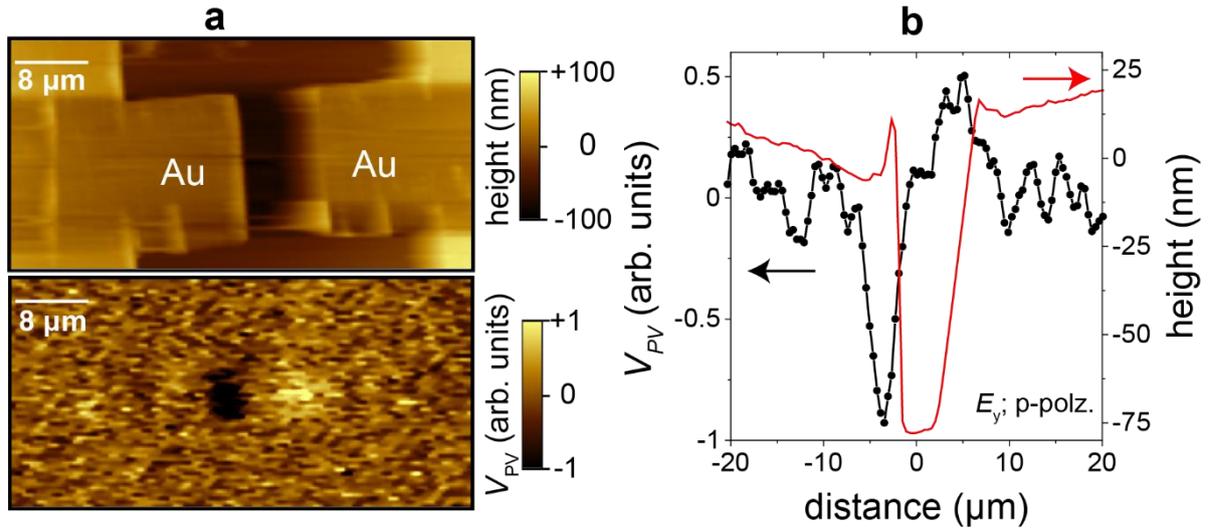

**Fig.5 | Mid-IR near-field photovoltaic response using a free electron laser.** (a) Topography (top) and PV response (bottom) maps obtained while scanning the near-field light across the graphene surface. (b) Horizontal line profiles of height (red) and PV signal (black) obtained at the center of the device over a 40 µm length.

## Conclusion

We developed a metal-insulator-metal campanile-shaped diamond probe that achieves ≈80% coupling of free-space mid-IR light into a ≈1 µm focal spot via non-resonant broadband adiabatic light compression, with minimal loss and full compatibility with high-power pulsed sources. When integrated into a scanning photovoltage microscope, it maps polarization and contact sensitive photocurrents in graphene with sub-wavelength resolution (≈1 µm). By tracking locally the signal change versus laser power and gate voltage, we isolate the photo-thermoelectric mechanism contribution, which in this case arises from localized, non-equilibrium carrier heating rather than bulk thermal effects, highlighting the unique sensitivity of mid-IR excitation to microscopic junction environments. Operation with a free-electron laser extends this capability to extreme mid-IR excitation regimes, establishing a robust, spectrally tunable platform that redefines the limits of nanoscale optoelectronic imaging in 2D materials[40-43]. This novel diamond based MIM campanile probe, with its engineered apex and broadband efficiency, sets a benchmark for mid-IR nanophotonics and provides a powerful route to disentangling carrier dynamics at unprecedented spatial and temporal scales and photoresponse in emerging quantum materials.

## Methods

### Experimental setup and methodology

Experiments were conducted in both far-field and near-field configurations using a quantum cascade laser (QCL) and free electron laser (FEL). For far-field measurements, the laser beam was directly focused onto the sample surface using a ZnSe lens (NA = 0.25), achieving a spot size of ≈30 µm. In contrast, for near-field measurements, the beam was focused (using the same ZnSe lens) onto the front surface of the MIM campanile pyramid, which transformed the far-field illumination into a near-field spot at its apex via the adiabatic compression process, enabling sub-wavelength spatial resolution as explained in the main text. The QCL provided light pulses at a repetition rate of 200 kHz, with a duration of 1 ns and a tunable wavelength range of $\lambda$ = 8.5 to 10.5 µm. The QCL laser beam was modulated at 500 Hz, facilitating lock-in detection of both PV responses and reflected light signals. Reflected light intensity signals were measured using a hybrid pyroelectric sensor featuring a $LiTaO_3$ pyroelectric element (GENTEC-AO). The electronic circuit employed for detecting the PV response is illustrated in Fig.2b. The PV response was measured using a variable gain custom high speed transimpedance amplifier. For accurate control of the surface approach of the MIM pyramid apex at the sample surface, the campanile pyramid is attached to one of the tines of a $LiNbO_3$ tuning fork[44]. The change in amplitude of the tine's oscillations at the tuning fork's resonance, is used as feedback for the sample-pyramid separation (Supplementary Fig. 8).

FEL measurements were performed at the FELIX facility in The Netherlands. The FEL delivers ≈8 µs long macropulses at a rate of 5, 10 or 20 Hz. Each macropulse consists of Fourier transform-limited spectrum micropulses (with typical duration of several ps long) with a repetition rate of 25, 50, or 1000 MHz, and the central wavelength tunable over a broad spectral range of 3 and 150 µm (Supplementary Fig. 8). For our experiments, wavelengths between 10 and 15 µm were utilized. Focusing the FEL light onto the sample surface was carried out similarly to the QCL experiments described above. The energy of the FEL macropulses was set to 1 mJ. To accurately measure the PV response due to FEL irradiation, a boxcar averager (UHFLI, Zurich Instruments) was used, effectively rejecting noise contributions outside the macropulse window. The electric components of the incoming light, $E_x$ and $E_y$, represent s and p-polarized light, respectively, along the x and y axes. To precisely control the excitation laser polarization and power, a combination of waveplate and a polarizer were used. Note that for the $\lambda$ dependence measurements done using QCL, the power of the laser at the sample plane was set to the maximum of the power delivered from a Hedgehog laser.

**FDTD Simulations**
We use commercial FDTD simulation software (Ansys Lumerical). In the metal-coated dielectric campanile probe, light coupling and confinement arise through a hybrid mechanism that combines waveguide mode transformation and plasmon-assisted adiabatic compression. The diamond core acts as a low-loss dielectric waveguide, enabling efficient far-field coupling at the probe base, while the gold coating introduces metal dielectric interfaces that support surface plasmon polaritons (SPPs). The interplay between these two mechanisms allows electromagnetic energy to propagate down the taper while being gradually compressed into a sub-wavelength region at the apex. In contrast to apertureless s-SNOM **or** aperture SNOM, where field enhancement primarily results from near-field scattering or dipolar antenna effects, the campanile geometry supports a continuously evolving guided mode. In s-SNOM, localized field enhancement occurs only at the metallic tip apex due to the excitation of localized surface plasmons, which depend strongly on the tip curvature and incident polarization. However, the energy coupling efficiency is limited by the abrupt transition between free-space and the tip's near field. The campanile probe, by contrast, exploits a tapered dielectric waveguide that ensures adiabatic mode compression, a gradual impedance matching between the optical mode at the base and the nanoscale mode at the apex. This adiabatic evolution minimizes back reflection and radiative loss, allowing the optical field to remain bound to the probe until its geometric dimensions approach the nanometric scale.

The local propagation constant along the taper, $\beta(z)$, governs the evolution of the optical mode as it propagates toward the apex. When the taper dimension becomes comparable to the wavelength (or $\lambda \approx a$, see Fig. 1(c) the effective refractive index, $n_{eff}(z)$ of the optical mode at position z along the taper increases due to stronger plasmonic coupling at the metal/dielectric interfaces. This enhanced coupling facilitates adiabatic energy compression of the mode with minimal reflection or scattering losses. As the effective cross-section area (A(z)) gradually decreases, the optical energy density changes as $U_E \propto |E|^2/A(z)$, leading to a sharp confinement of the electromagnetic field near the apex. To quantify this, we analyzed the area-averaged field intensity enhancement, defined as $\frac{<|E|^2>}{|E_0|^2} = KA^m$, where K is a proportional constant, and m characterizes how the enhancement scales with the focusing mode (apex-)area. The data shown in Fig. S2(d) were fitted using the power-law relation, yielding K = 340 and m ≈ -1 indicates that the area-averaged intensity enhancement is inversely proportional to the mode area, consistent with energy flux conservation under adiabatic compression.

The apex angle of the taper plays a crucial role in determining adiabaticity and focusing efficiency. A smaller apex angle yields a slower spatial variation of $\beta(z)$, better satisfying the adiabatic condition and thus enabling more efficient mode transformation and field localization. Conversely, a large angle causes rapid changes in the propagation constant, increasing reflection and mode mismatch, thereby reducing the focusing efficiency. Therefore, the enhanced field localization observed for smaller apex angles can be attributed to improved impedance matching and stronger plasmon-dielectric hybrid confinement. However, for our experiments we have used campanile probes with θ = $80^0$.

**Sample preparation.** Single crystal CVD graphene was grown deterministically on Cu-foil and transferred onto $SiO_2$/Si substrate (285 nm of $SiO_2$) via semi-dry transfer[45]. Two-terminal device fabrication was carried out using electron beam lithography followed by reactive ion etching to define the device channels and thermal evaporation of Au/Ni electrodes (60/7 nm). Polymer remover (AR 600-71) was utilized after each fabrication step to ensure an ultra-clean graphene surface[46]. High quality GaSe flakes were exfoliated from a Bridgman-grown GaSe bulk crystal and dry transferred onto the graphene, capping the graphene channel. In the Au/graphene/Au devices, the two Au electrodes serve as source (*s*) and drain (*d*) to a square graphene channel of length *l* = 10 μm.

**Acknowledgements**

This work is supported by UKRI Research Grant High Performance Wide spectral range Nanoprobe (HiWiN) EP/V00787X/1, by the Engineering and Physical Sciences Research Council (Grant Nos. EP/X020304/1, EP/T019018/1 and EP/X524967/1), EU Graphene Flagship Core3 project, and the Defence Science and Technology Laboratory (DSTL). We express our gratitude to HFML-FELIX and Ben Murdin for providing us with the lab space required for FEL based experiments and hosting the HiWiN as a user facility at FELIX, open for cutting-edge research in IR through THz spectroscopy for physics and chemistry. The authors specifically thank Britta Redlich (currently The Director, DESY) and Peter Weightman of Liverpool University for generating and supporting the idea of combining FEL and SPM, as well as and Andrei Kirilyuk (FELIX, Radboud University), Matteo Savoini (ETH Zurich, Switzerland) and Sam Khorsand (ASML, The Netherlands) for stimulating discussions. We acknowledge excellent engineering of diamond campanile probes by Loek van den Boom at Dutch Diamond Technologies, The Netherlands, and scanning probe microscopy integration supported by expert technical advice from Peter De Wolf and Hartmut Stadler from Bruker Ltd. The authors acknowledge Zakhar Kudrynskyi (University of Nottingham) for the GaSe bulk crystals.

**Author contributions**
O.V.K. and B.R. initiated and conceived the project, O.V.K. and R.M. designed the diamond campanile tetragonal pyramid and integrated it into Bruker Innova microscope. R.M. built the measurement set-up in both far-field and near-field configurations and performed the metallization of diamond campanile probes. R.M. and K.A. performed and analyzed the FDTD simulations. O.V.K. designed electronics for low noise and mid-IR detection using pyro electric detectors. N.D.C., B.T.D. and A.P. characterized the graphene and GaSe/graphene devices. R.M. N.D.C., B.T.D., and N.D. measured the photovoltaic responses of the devices using both quantum cascade laser and FEL. R.M. drafted the initial manuscript, with all co-authors contributing to the final version. O.V.K supervised the project.

# Supplementary Information for

# Sub-wavelength mid-infrared imaging of locally driven photocurrents using diamond campanile probes

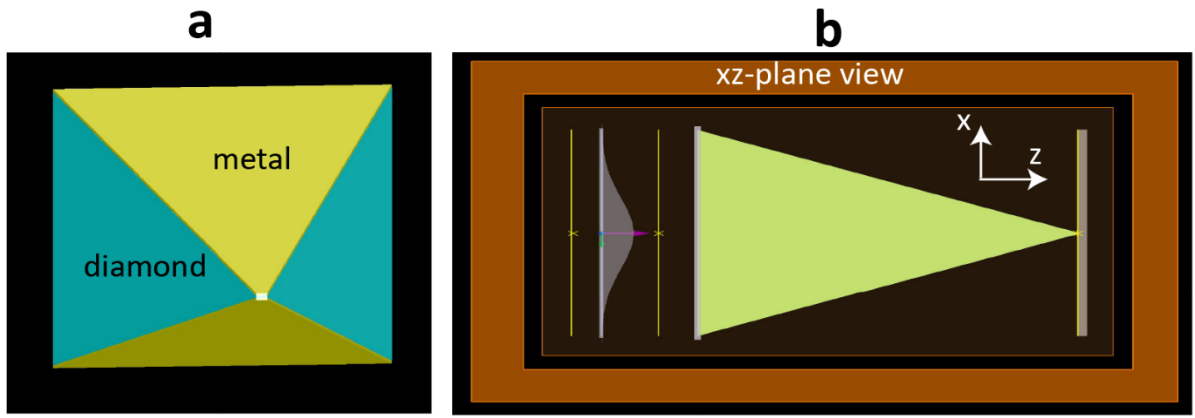

**Supplementary Figure 1. FDTD simulation setup.** (a) A schematic diagram of the metal/diamond/metal (metal/insulator/metal; MIM) campanile pyramid structure used in the simulation. (b) An xz plane view of the simulation scheme is shown. Yellow vertical lines represent the 2D monitors positioned to collect the field profiles at various locations within the simulation frame. Additionally, a point monitor is placed a few nanometers away from the centre at the apex to measure localized field intensities. The purple arrows indicate the propagation direction of the light pulse (gray Gaussian pulse) incident. The base of the pyramid has an opening of 15 μm x 15 μm while the apex dimensions were varied. In order to vary the value of $θ$, the distance between the base and apex was varied.

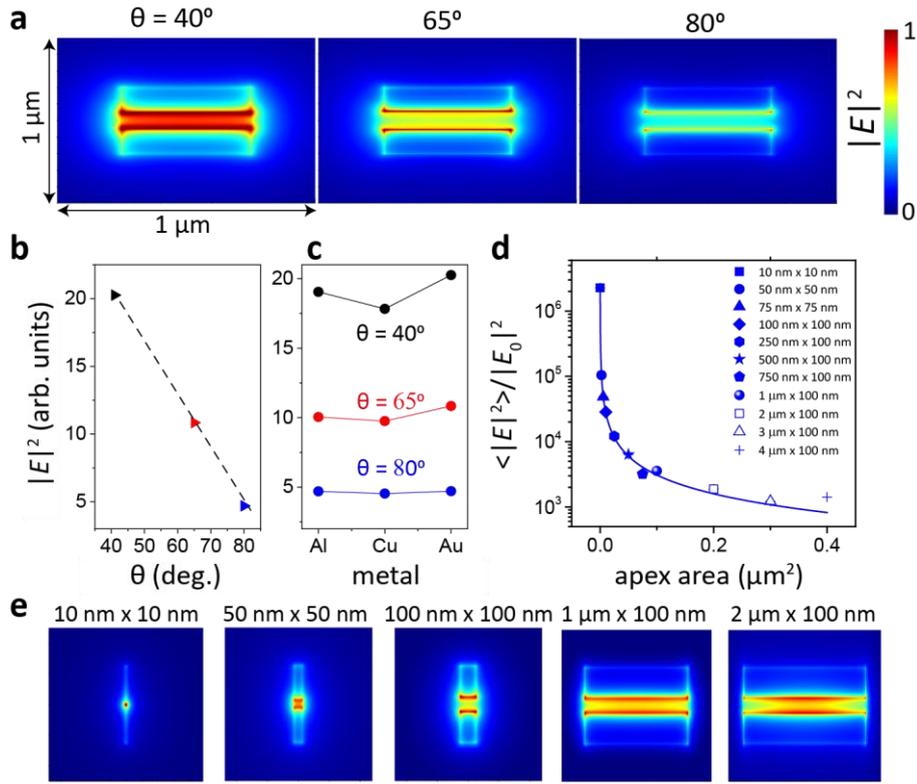

**Supplementary Figure 2. FDTD simulation - mid IR near field *E*-field distribution and intensity analysis at the apex of the MIM campanile pyramid.** (a) The xy-plane view of the *E*-field intensity distribution at the apex of MIM campanile pyramid is shown for three different apex angles; $\theta = 40^0$, $65^0$, and $80^0$. These cases illustrate the variation in the adiabatic light spot's efficiency depending on the apex angle, with a clear enhancement of field confinement as $\theta$ decreases. (b) The integrated *E*-field intensity at a pixel located at the origin (apex center) of images is plotted against $\theta$. (d) Similarly, the integrated *E*-field intensity at the apex is shown for three different elements (aluminium, copper, and gold) used to metallize the pyramid. (e) Area-averaged intensity enhancement, $<|E|^2>/|E_0|^2$ as a function of apex area. Here, $|E_0|$ is the amplitude of the incident *E*-field. The blue solid curve represents a power-law fit to the simulated data. Different symbols correspond to different (x, y) sizes of the apex used in the simulations. (e) The *E*-field intensity $|E|^2$ showing the field localization at the apex plane for five different apex dimensions (x, y).

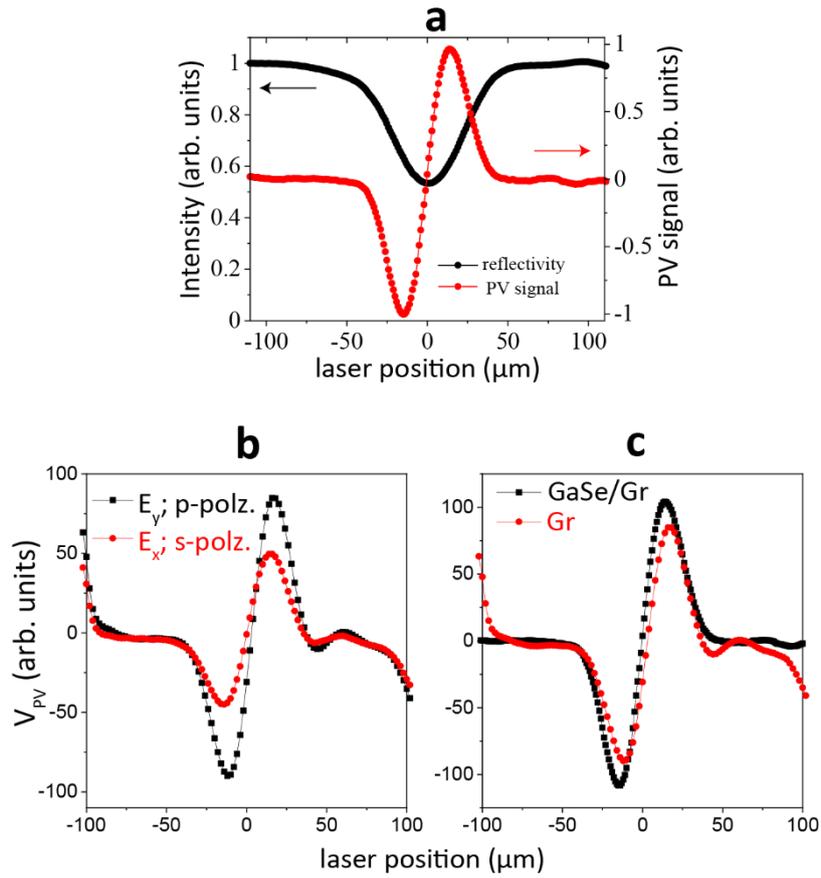

**Supplementary Figure 3. Far-field measurements: Polarization and sample stack dependence.** The horizontal line profiles, corresponding to the reflectivity and $V_{PV}$ response, taken at the center of the images shown in Fig. 3(a) (black) and 3(b) (red) of the main text are plotted. (a) Horizontal line profiles of the $V_{PV}$ signal measured in a graphene-device for two orthogonal excitation laser polarizations. (b) same as (a) but measured in two different samples GaSe/graphene (black) and graphene (red) devices for a given excitation laser polarization of $E_y$, showing a nearly identical trend.

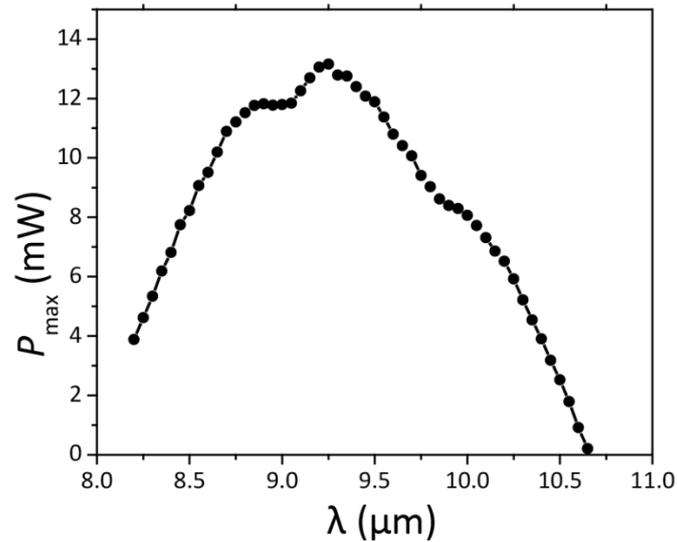

**Supplementary Figure 4. Maximum laser power as a function of operating wavelength**. Maximum laser power at sample plane derived from the QCL is plotted against $\lambda$.

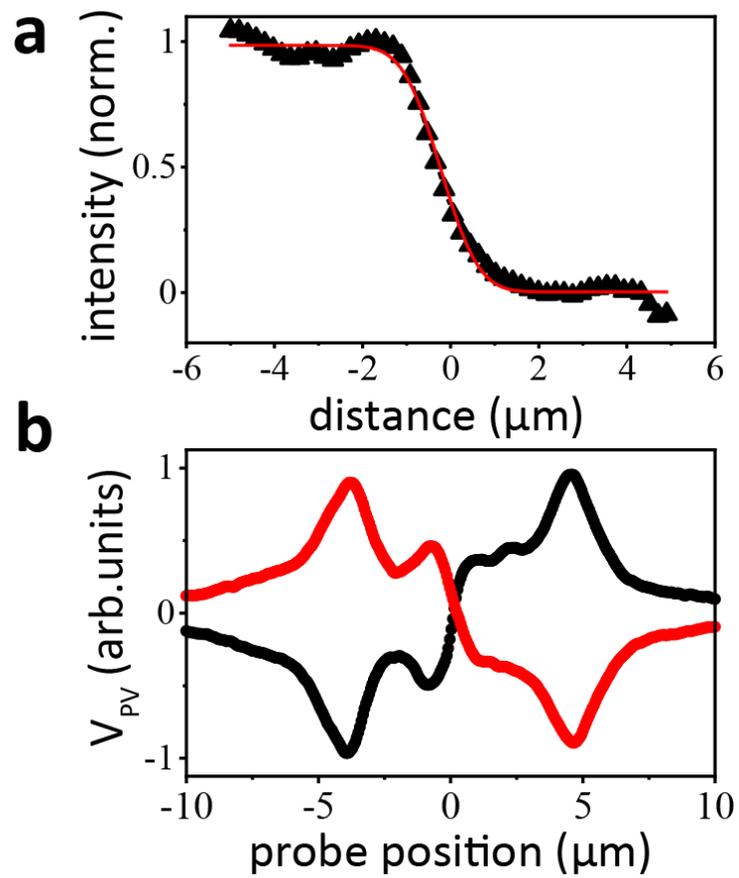

**Supplementary Figure 5. Near-field measurements - adiabatic spot calibration and near-field response of GaSe/graphene device.** (a) Line profile of normalized reflected light intensity along the border separating the gold contact and sample region over a 10 µm horizontal length (shown in Fig. 4b of main text). (b) Line profile of the $V_{PV}$ response measured as the transformer is scanned across the centre of the device for a horizontal distance of 20 µm (see Fig. 4c of the main manuscript). The $V_{PV}$ response is also shown for the case where the polarity of the Au contact terminals is inverted.

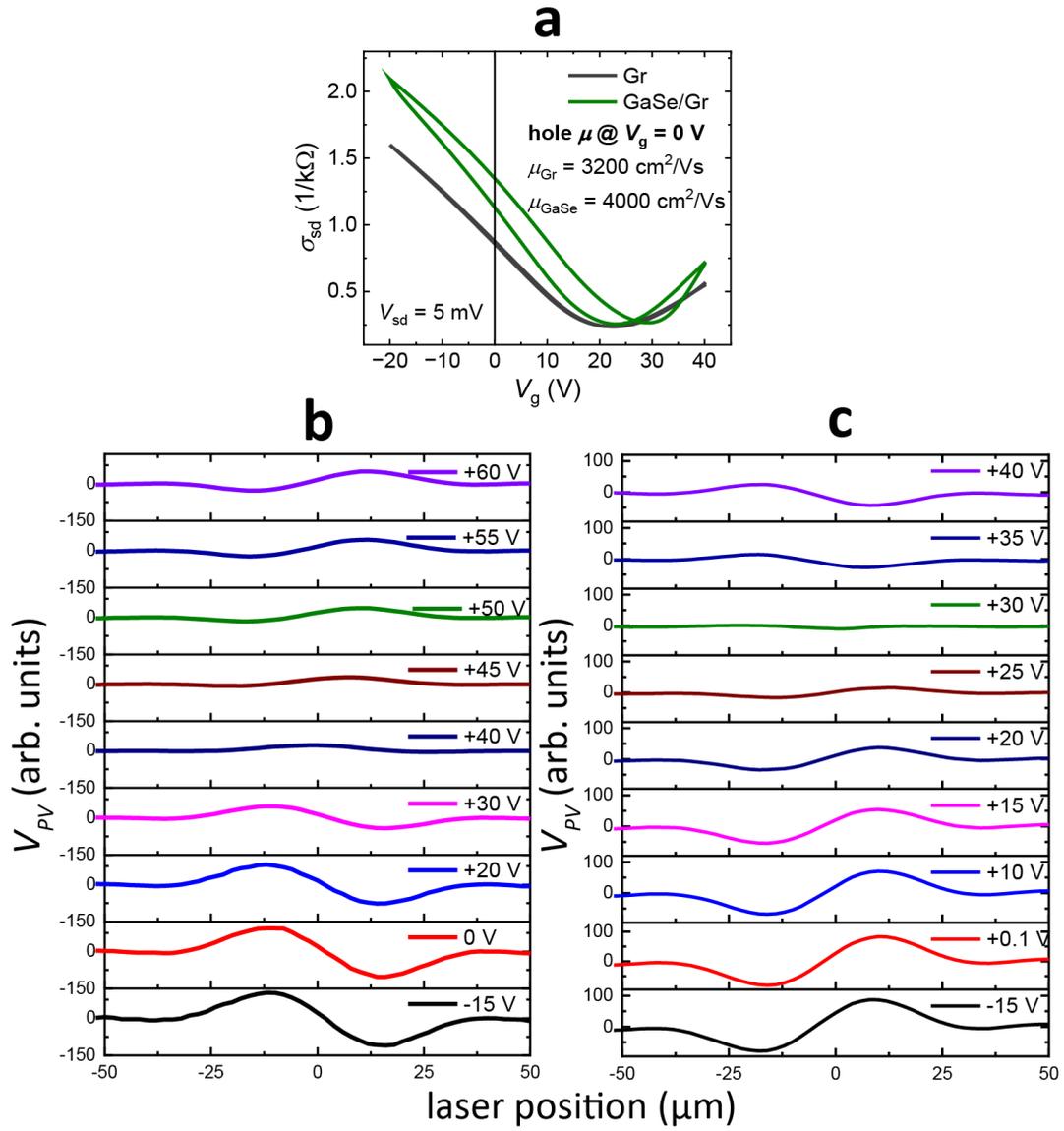

**Supplementary Figure 6. Far-field measurements: back-gate voltage dependence**. Source-drain conductivity, $\sigma_{Sd}$ as a function of back-gate voltage, $V_g$ for graphene (black) and GaSe/graphene (green) devices, showing the location of their respective Dirac points. (b,c) Horizontal line profiles of the $V_{PV}$ signal measured in the graphene device (b) and GaSe/graphene device (c) for various $V_g$ values swept across the Dirac point. Note the polarity reversal of the peak $V_{PV}$ signal above and below the Dirac point in both cases.

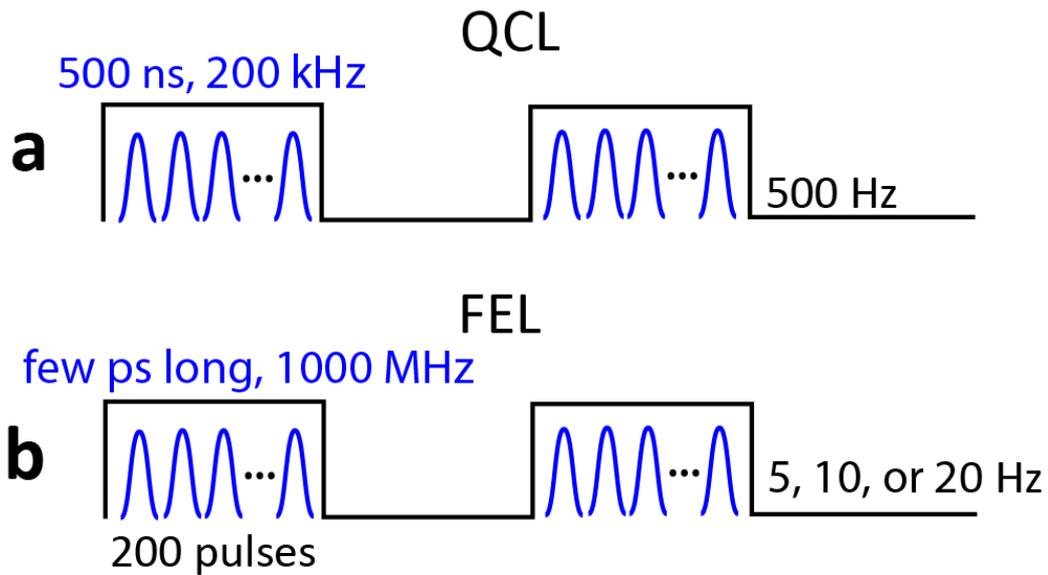

**Supplementary Figure 7. Excitation laser parameters**. Illustration showing the train of laser pulses derived from QCL (a) and FELIX (b).

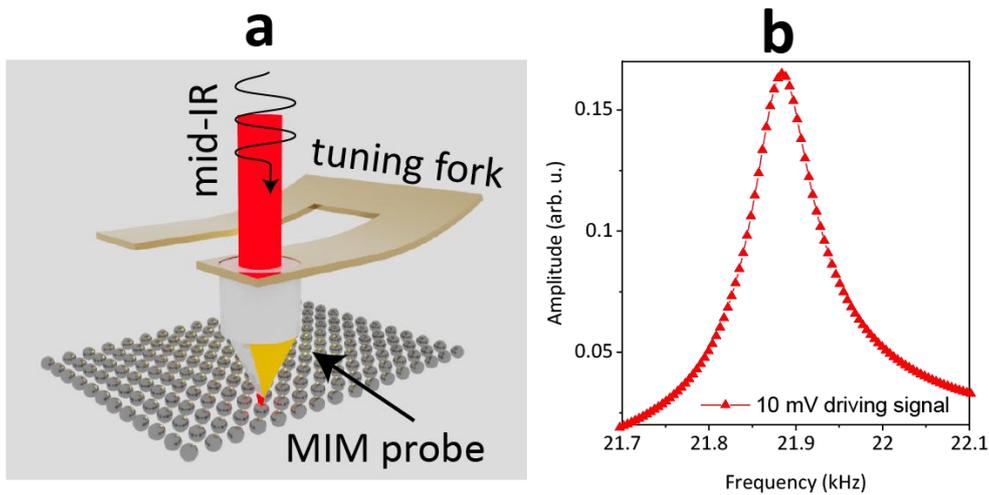

**Supplementary Figure 8. Illustration of tuning fork resonance mechanism for SPM feedback.** (a) The MIM campanile pyramid is mounted on one of the tines of a $LiNbO_3$ tuning fork, which serves as a sensitive probe for maintaining feedback while approaching sample surface. This configuration enables precise control of the distance between the apex of the pyramid and the sample surface, essential for achieving accurate measurements and ensuring stability during near-field measurements. (b) The resonance behaviour of the $LiNbO_3$ tuning fork is shown as driven by the internal lock-in amplifier of Bruker Innova SPM. The amplitude of the TF oscillations near its resonance is used as feedback.